\documentclass[aps,pra,twocolumn,amsfonts,amsmath,showpacs,floatfix,footinbib,groupedaddress,superscriptaddress]{revtex4}
\usepackage{amssymb,graphics,graphicx,times,bm,bbm,multirow}

\begin{document}

\title{Equation of motion for process matrix: Hamiltonian identification and dynamical control of open quantum systems}
\author{M. Mohseni}
\affiliation{Research Laboratory of Electronics, Massachusetts
Institute of Technology, 77 Massachusetts Ave., Cambridge, MA 02139,
USA}
\author{A. T. Rezakhani}
\affiliation{Department of Chemistry and Center for Quantum Information Science and Technology, University of Southern
California, Los Angeles, CA 90089, USA}

\begin{abstract}
We develop a general approach for monitoring and controlling
evolution of open quantum systems. In contrast to the master
equations describing time evolution of density operators, here, we
formulate a dynamical equation for the evolution of the process matrix
acting on a system. This equation is applicable to non-Markovian
and/or strong coupling regimes. We propose two distinct applications
for this dynamical equation. We first demonstrate identification of
quantum Hamiltonians generating dynamics of closed or open systems
via performing process tomography. In particular, we argue how one
can efficiently estimate certain classes of sparse Hamiltonians by
performing partial tomography schemes. In addition, we introduce a
novel optimal control theoretic setting for manipulating quantum
dynamics of Hamiltonian systems, specifically for the task of
decoherence suppression.
\end{abstract}

 \maketitle

\textit{Introduction.}---Characterization and control of quantum systems are among the most
fundamental primitives in quantum physics and chemistry
\cite{nielsen-book,qcontrol}. In particular, it is of paramount
importance to identify and manipulate Hamiltonian systems which have
unknown interactions with their embedding environment \cite{kosut}.
In the past decade, several methods have been developed for
estimation of quantum dynamical processes within the context of
quantum computation and quantum control
\cite{nielsen-book,aapt,dcqd1,dcqd2,Emerson,bendersky-paz}. These
techniques are known as ``quantum process tomography" (QPT), and
originally were developed to estimate the parameters of a
``superoperator" or ``process matrix", which contains all
information about the dynamics. This is usually achieved through an
inversion of experimental data obtained from a complete set of state
tomographies. QPT schemes are generally inefficient, since for a
complete process estimation the number of required experimental
configurations and the amount of classical information processing
grows exponentially with size of the system. Recently,
alternative schemes for partial and efficient estimation of quantum
maps have been developed
\cite{dcqd1,dcqd2,Emerson,bendersky-paz,dcqd3,HI} including
efficient data processing for selective diagonal \cite{Emerson} and
off-diagonal parameters of a process matrix \cite{bendersky-paz}.
However, it is not clear how the estimated elements of the
process matrix could help us actually characterize the set of
parameters for Hamiltonians generating such dynamics. These
parameters of interest generally include the system free
Hamiltonians and those coupling strengths of system-bath
Hamiltonians. More importantly, it is not fully understood how the
relevant information obtained from quantum process estimation
experiments can be utilized for other applications such as optimal
control of a quantum device.

In this work, we develop a novel theoretical framework for studying
general dynamics of open quantum systems. In contrast to the usual
approach of utilizing master equations for density operator of a
quantum system, we introduce an equation of motion for the evolution
of a process matrix acting on states of a system. This equation does
not presume Markovian or perturbative assumptions, hence provides a
broad approach for analysis of quantum processes. We argue that the
application of partial quantum estimation schemes
\cite{dcqd1,dcqd2,Emerson,bendersky-paz,dcqd3,HI} enables efficient
estimation of sparse Hamiltonians.
Furthermore, the dynamical equation for process matrices leads to
alternative ways for controlling generic quantum Hamiltonian
systems. In other words, one can utilize this equation to drive the
dynamics of a(n) closed (open) quantum system to any desired target
quantum operation. In particular, we apply quantum control
theory to find the optimal fields to decouple a system from its
environment, hence, ``controlling decoherence''.

\textit{Dynamical equation for open quantum systems.}---In quantum theory, the evolution of a system---assuming separable
initial state of the system and environment---can be described by a
(completely-positive) quantum map $\mathcal{E}_{t}(\varrho
)=\sum_{i}A_{i}(t)\varrho A_{i}^{\dagger }(t)$, where $\varrho $ is
the initial state of the system \cite{open-book}. An alternative,
more useful expression is obtained by expanding
$A_{i}(t)=\sum_{m}a_{im}(t)\sigma _{m}$ in $\{\sigma
_{k};k=0,1,\ldots ,d^{2}-1\}$ (a fixed operator basis for the
$d$-dimensional Hilbert space of the system) which leads to
$\mathcal{E}_{t}(\varrho )=\sum_{mn=0}^{d^{2}-1}\chi
_{mn}(t)\sigma _{m}\varrho \sigma _{n}^{\dagger }$. The positive-Hermitian process matrix $\bm{\chi}%
(t)=\bigl[\sum_ia_{im}(t)\bar{a}_{in}(t)\bigr]$ represents $\mathcal{E}_{t}$ in the $\{\sigma _{k}\}$
basis, where bar denotes complex conjugation. The process matrix elements $%
\chi _{mn}(t),$ in any specific time $t$, can be experimentally
measured by any QPT scheme \cite{experiment}.


In an open quantum system, the time-dependent Hamiltonian of the
total system-bath ($SB$) has the general form $H(t)=H_{S}(t)+H_{B}(t)+H_{SB}(t)$, where $S$ ($B$) stands for the system (bath or the surrounding environment). We denote
the evolution operator which is generated from this Hamiltonian, from
time $0$ to $t$, by $U(t)$. The
Hamiltonian $H_{SB}(t)$ can be written as $H_{SB}(t)=\sum_{k}\lambda _{k}(t)\sigma _{k}\otimes
B_{k}$, where $\lambda _{k}(t)$s are the coupling strengths of the
system-bath interaction, and $\{B_{k}\}$ are some bath operators.
Now we describe the dynamics in the interaction picture by
introducing the time evolution operators
$U_{0}(t)$, $U_S(t)$, and $U_B(t)$, generated by $H_{0}=H_{S}\otimes
I_{B}+I_{S}\otimes H_{B}$, $H_S$, and $H_B$, respectively.

The system-bath Hamiltonian in the interaction picture becomes:
$H_{I}(t)=U_{0}^{\dagger }(t)H_{SB}(t)U_{0}(t)$. By
introducing $\widetilde{\sigma }_{k}(t)=U_S^{\dag}(t)\sigma
_{k}U_S(t)\equiv \sum_{l}s_{kl}(t)\sigma _{l}$ and
$\widetilde{B}_{k}(t)=U_B^{\dag}(t)B_{k}U_B(t)$
as the rotating operators under the evolution of the
free Hamiltonian of the system and bath, we can rewrite
$H_{I}(t)=\sum_{k}\lambda _{k}(t)\widetilde{\sigma }_{k}(t)\otimes \widetilde{%
B}_{k}(t)$. The Schrodinger equation in the interaction picture can
be expressed as:

\begin{eqnarray}
i\text{d}A_{i}^{I}(t)/\text{d}t=\sum_{k}H_{ik}^{\prime
}(t)A_{k}^{I}(t),
\end{eqnarray}
where $U_{I}(t)=U_{0}^{\dagger }(t)U(t)$, $A_{i}^{I}(t)=~_{B}\langle
b_{i}|U_{I}(t)|b_{0}\rangle _{B}\equiv \sum_{m}a_{im}^{I}(t)\sigma
_{m}$ are the Kraus operators at the interaction picture,
$H'_{ij}(t)=\sum_{pq}\lambda _{p}s_{pq}~_{B} \langle
b_{i}|\widetilde{B}_{p}|b_{j}\rangle _{B}\sigma _{q}$, and
$\{|b_{i}\rangle \}$ is a basis for the bath Hilbert space. The
interaction picture $\bm{\chi}$ matrix is defined as $\chi
_{mn}^{I}(t)=\sum_{i}a_{im}^{I}(t)\bar{a}_{in}^{I}(t)$, which is
related to the elements of the measured process matrix through
$\chi^I_{mn}(t) =
\sum_{m'n'}\chi_{m'n'}(t)\text{Tr}[\sigma_mU^{\dag}_S(t)\sigma_{m'}]
\text{Tr}[\sigma_n U_S^T(t)\sigma_{n'}].$ Thus, the time evolution
of the $a_{m}^{I}$ coefficients reads as:
\begin{eqnarray}
&i\text{d}a_{im}^{I}/\text{d}t=\sum_{klpq}a_{kl}^{I}\lambda _{p}s_{pq}\alpha
_{~~m}^{qp}~_{B}\langle b_{i}|\widetilde{B}_{p}|b_{k}\rangle _{B},
\label{aI-dot}
\end{eqnarray}
where $\alpha _{~~m}^{kl}=\text{Tr}[\sigma _{k}\sigma_{l}\sigma_m^{\dag}]$.
From this equation, one can obtain the time evolution of $\bm{\chi}^I$ as follows:
\begin{eqnarray}
&i \text{d}\bm{\chi}^{I}/ \text{d}t=\widetilde{\bm{H}}\bm{K}-\bm{K}^{\dag }\widetilde{\bm{H}}
^{\dag }, \label{open-EQ1}
\end{eqnarray}
where
\begin{eqnarray}
&[\widetilde{\bm{H}}]_{n(imj)}=&\textstyle{\sum_{pq}}\lambda
_{p}s_{pq}\alpha
_{~~n}^{qp}~_{B}\langle b_{j}|\widetilde{B}_{m}|b_{i}\rangle _{B}, \label{p-H}\\
&[\bm{K}]_{(imj)n}=&a_{im}^{I}\bar{a}_{jn}^{I},  \label{K}
\end{eqnarray}%
in which $(imj)$ is considered as a new single index. The order of
the pseudo-Hamiltonian $\widetilde{\bm{H}}$ is $d^{2}\times d^{6}$, but number of independent parameters is $\leqslant d^2$, which is the
maximum number of nonzero $\lambda _{p}$s.  By using a generalized
commutator notation $\left[ A,B\right] ^{\star }\equiv AB-B^{\dag}A^{\dag }$, Eq.~(\ref{open-EQ1}) can be represented in the following form:
\begin{eqnarray}
&i\mathrm{d}\bm{\chi}^{I}/\mathrm{d}t=[\widetilde{\bm{H}},\bm{K}]^{\star }.
\label{open-EQ}
\end{eqnarray}
This is the (super-) dynamical equation for open quantum systems, i.e.,
an equation for the time variation of quantum dynamics itself, in
which no state of the system appears, in contrast to the existing master equations \cite{open-book}.

The knowledge of the $\bm{K}$ matrix is generally required for application of Eq.~(\ref{open-EQ}).
The $\bm{\chi}^{I}$ matrix can be diagonalized by the
unitary operator $\bm{V} $: $\bm{\chi}^{I}=\bm{V}\bm{D}\bm{V}^{\dag
}$, where $\bm{D}=\text{diag}(D_i)$. Then, the Kraus operators in the interaction picture are
$A_{i}^{I}(t)=\sqrt{D_{i}}\sum_{m}V_{mi}\sigma _{m}$ \cite{nielsen-book}. Hence, we obtain $a_{im}^{I}=\sqrt{D_{i}}V_{mi}$ and $K_{imjn}=\sqrt{D_{i}D_{j}}V_{mi}%
\overline{V}_{nj}$. Diagonalization of a sparse $\bm{\chi}^{I}$ matrix,
hence construction of the $\bm{K}$ matrix, can be performed
efficiently. The unknown parameters of Eq.~(\ref{open-EQ}) are
elements of $\widetilde{\bm{H}}$ matrix which contain the
information about the system-bath coupling strengths $\lambda _{k}$.

For unitary evolutions, following a similar approach, the dynamical equation for the process matrix is obtained as:
\begin{eqnarray}
&i\mathrm{d} \bm{\chi}/\mathrm{d}t=[\widetilde{\bm{H}},\bm{\chi}]^{\star },
\label{closed-EQ}
\end{eqnarray}
where $\bm{\widetilde{H}}=[\widetilde{h}_{ml}]$, $\widetilde{h}_{mn}(t)\equiv
\sum_{k}\alpha _{~~m}^{kn}h_{k}(t)$, and $h_l(t)$ are defined
through $H(t)=\sum_{l}h_{l}(t)\sigma_{l}$. It should be noted that Hermiticity of
$H$ implies only $d^2$ real independent parameters in $\widetilde{\bm{H}}$, which can
be estimated via QPT schemes.

\textit{Hamiltonian identification.}---Consider a large ensemble of the identically prepared systems in the
state $\varrho $, half of which are measured after duration $t$, and
the rest are measured at $t+\Delta t$, where $\Delta t$ is small
relative to $t$. Thus, by performing any type of QPT strategy one can
obtain the matrix elements $\chi_{mn}(t)$ and $\chi _{mn}(t+\Delta
t)$, hence their derivatives $\dot{\chi}_{mn}(t)\approx\left( \chi
_{mn}(t+\Delta t)-\chi _{mn}(t)\right) /\Delta t$ with accuracy
$\Delta t$. Consequently, using Eq.~(\ref{closed-EQ})
(Eq.~(\ref{open-EQ})) one can in principle identify the free
[system-bath] Hamiltonians for closed (open) quantum systems.

For unitary evolutions, a simple relation between the elements of
the $\bm{\chi}$ matrix and the system Hamiltonian parameters is
obtained, up to the second order in $t$ and a global phase
$\text{Tr}[H]$, as:
$\chi
_{00}(t)\approx\ 1-\frac{1}{2}t^2\sum_{ij}\alpha^{ij}_{~~0}h_ih_j+ \bar{\alpha}%
^{ij}_{~~0}\bar{h}_i\bar{h}_j$, $\chi
_{m0}(t)\approx\-ih_mt-\frac{1}{2}t^2\sum_{ij}\alpha^{ij}_{~~m}h_ih_j$,
and
\begin{eqnarray}
\chi _{mn}(t)\approx\ h_m\bar{h}_nt^2, \label{ShortTimeChi}
\end{eqnarray}
where $m,n\neq0$. From Eq.~(\ref{ShortTimeChi}), for a given short time $%
t$, we have $h_{n}=e^{i\varphi _{n}}\sqrt{\chi _{nn}}/t$, from which the relative errors satisfy $\text{Re}%
[\delta h_{n}/h_{n}]=\delta \chi _{nn}/2\chi _{nn}$. According to
the Chernoff bound \cite{dcqd3,chernoff}, to estimate $\chi _{nn}$s with
accuracy $\Delta\geqslant|\chi _{nn}-\overline{\chi }_{nn}|=\delta
\chi _{nn} $ --- where $\overline{\chi}_{nn}$ is the average of the
results of $M$ repeated measurements --- with success probability
greater than $1-\epsilon $, one needs $M=\mathcal{O}(|\log
\epsilon/2|/\Delta^2)$. Information of the phases $\varphi _{n}$, up
to a global phase, can be estimated by measuring $\chi_{mn}$s for
$m\neq n$.

Using the above construction, next we discuss efficient Hamiltonian
identification schemes via performing certain short-time scale QPTs.
A precursor to this type of short-time expansion in order to
efficiently obtain process matrix parameters can be found in
Ref.~\cite{levi}, however, its underlying models, the assumptions,
and the identification method are more restrictive and generally
incommensurable with ours.

In generic $N$-body physical systems (e.g., $N$ qubits),
interactions are $L$-local where $L$ is typically $2$. That is,
$H=\sum_{k}H_{k}$, where each $H_{k}$ includes only interactions of
$L$ subsystems, with overall $\mathcal{O}(N^{L})$ independent
parameters. This implies that in the $\{\sigma _{k}\}$ basis $H$ has
a sparse-matrix representation. Hence, the number of free parameters
of the corresponding unitary or process matrix, unlike their
exponential size, will be polylog($d$) (i.e., a polynomial of $N$).
Here, we mainly concentrate on controllable $L$-local Hamiltonians,
which are of particular interest in the context of quantum
information processing in order to generate a desired quantum operation. An
important example of this class is the Heisenberg exchange
Hamiltonian in a network of spins with nearest neighbor
interactions. This 2-local sparse Hamiltonian (in the Pauli basis)
also generates a sparse process matrix \cite{HI} and is
computationally universal over a subspace of fixed angular momentum
\cite{heisenberg}.

Let us consider a sparse Hamiltonian,
$H(t)=\sum_{m}h_{m}(t)\sigma_{m}$, with polylog$(d)$ nonzero $h_m$s,
where $\{\sigma_m\}$ is the nice error basis \cite{dcqd2}. In the
short-time limit, according to Eq.~(\ref{ShortTimeChi}), if the
Hamiltonian is sparse in the $\{\sigma _{k}\}$ basis, only
polylog$(d)$ of $h_{m}$s would be nonzero. Thus, number of nonzero elements in the $\bigl[t^{2}h_{m}\bar{h}_{n}\bigr]_{m,n\neq 0}$ block would be also of the same
order. \textit{A priori} knowledge of the general form of a given
sparse Hamiltonian leads to [up to $\mathcal{O}(t^{3})$] nonzero elements in the $\bigl[t^{2}h_{m}\bar{h}_{n}\bigr]_{m,n\neq 0}$ block, according to Eq.~(\ref{ShortTimeChi}).
Therefore, if we can efficiently determine all nonzero elements
of this block, we would have polylog$(d)$ quadratic equations from
which we can estimate $h_{m}$s. In other words, by only
\textit{polynomial} experimental settings we would be able to
extract relevant information about the Hamiltonian from a suitable QPT experiment \cite{experiment}. In general,
there are three distinct process estimation techniques, including
Standard Quantum Process Tomography (SQPT) \cite{nielsen-book},
Direct Characterization of Quantum Dynamics (DCQD) \cite{dcqd1}, and Selective Efficient Quantum Process Tomography (SEQPT) \cite{bendersky-paz}. The scaleup of physical resources varies among
these process estimation strategies \cite{dcqd3}. SQPT is
inefficient by construction, since we still need to measure an
exponentially large number of observables in order to reconstruct
the process matrix through a set of state tomographies. SEQPT can efficiently estimate quantum sparse
Hamiltonians via selectively estimating a polynomial number of
$\chi_{mn}$s associated to the Hamiltonian, within the context of
short-time analysis. Using the DCQD scheme, in short-time limit, one
can also efficiently estimate all the parameters of certain sparse
Hamiltonians, specifically all the diagonal elements $\chi_{nn}$ --- a detailed analysis thereof is beyond the discussion of this work and will be presented in another publication \cite{xx}. Note that in contrast to SQPT, both DCQD and SEQPT
assume access to noise-free ancilla channels. However, recently a
generalization of the DCQD scheme to certain cases of calibrated faulty
preparation, measurement, and auxiliary systems has been developed \cite{AIP}.

We emphasize that, within the context of short-time analysis, the
efficient estimation is only applicable to the Hamiltonians for
which the location of nonzero elements in a given basis is known
from general physical or engineering considerations, such as in the
exchange Hamiltonian in solid-state quantum information processing
\cite{heisenberg}. The exchange Hamiltonian describes the underlying
interactions for various systems, such as spin-coupled quantum dots
\cite{Loss:98}, donor-atom nuclear/electron spins
\cite{Kane:98Vrijen:00}, semiconductor quantum dots \cite{Petta05},
and superconducting flux qubits \cite{niskanen}. The anisotropic
exchange Hamiltonian exists in quantum Hall systems \cite{Hall},
quantum dots/atoms in cavities \cite{Imamoglu:99}, exciton-coupled
quantum dots \cite{exciton}, electrons in liquid-Helium
\cite{Platzman:99}, and neutral atoms in optical lattices
~\cite{opticallattices}.

\textit{Applications to quantum dynamical control.}---One immediate application of any equation of motion --- i.e, dynamical equation --- for a quantum
or classical system is to manipulate its state or dynamics toward a
desired target. The ability to control quantum dynamics by certain
external control fields is essential in many applications including
physical realizations of quantum information devices. Due to environmental noise and device imperfections, it is generally difficult
to maintain quantum coherence during dynamical evolution of quantum systems. Reducing or controlling decoherence, therefore, is an
important objective in a control theoretic investigation of quantum
systems.

Optimal control theory (OCT) \cite{oct}, has been developed for
finding control fields to guide a quantum system, subject to natural
or engineering constraints, as close as possible to a particular
target. For closed quantum system, OCT has been proposed for
controlling states \cite{oct} and unitary dynamics
\cite{palao-kosloff}. In OCT, a quantum system is driven from an
initial state or unitary operation to a final configuration, via
applying external fields. This is achieved, for example, by
modifying a free Hamiltonian $H_0$ as $H(t)=H_0-\mu\pi(t)$, where
$\mu$ is a system operator (e.g., atomic or molecular dipole moment)
and $\pi(t)$ is a shaped external field (e.g., laser pulse)
\cite{palao-kosloff}. The optimization is based on maximizing a
yield functional $\widetilde{\mathcal{Y}}$, e.g., fidelity of the
final and target configurations, by a variational procedure
($\delta\widetilde{\mathcal{Y}}/\delta\pi=0$) subject to a set of
constraints.

Having an equation of motion implies how one can control dynamics of
a system toward a target configuration. Thus, a new method for
controlling dynamics of \textit{open} quantum systems can be
developed by our equation of motion (Eq.~(\ref{open-EQ})),
specifically applicable to optimal decoherence control. For isolated
systems we have $\lambda_k=0$ (hence $\widetilde{\bm{H}}=0$), from
which one can obtain $\chi^I_{mn}=\delta_{m0}\delta_{n0}\equiv
[\bm{E}_{00}]_{mn}$. However, due to decoherence or other
environmental effects, there might be some residual interaction
$\widetilde{\bm{H}}_0$ between the system and environment. Our
objective, here, is to apply a control field $\pi(t)$ to modify the
pseudo-Hamiltonian, Eq.~(\ref{p-H}), in order to suppress the
decohering interaction. Since $\widetilde{\bm{H}}$ is linear in
$\lambda$s, applying a control coupling field would affect
$\widetilde{\bm{H}}$ linearly. Thus, if we introduce an external
controllable field $\pi(t)$, the pseudo-Hamiltonian
$\widetilde{\bm{H}}_0$ becomes
$\widetilde{\bm{H}}(t)=\widetilde{\bm{H}}_0 -\bm{\mu}\pi(t)$, where
$\bm{\mu}$ is a system operator. The control strategy is to find the
optimal $\pi(t)$ such that the constrained fidelity
\begin{eqnarray}
&\widetilde{\mathcal{Y}}=\text{Re}\bigl[\mathcal{Y}- \int_0^T \mathrm{d}t~ \text{Tr}\{( \mathrm{d}\bm{\chi}^I/ \mathrm{d}t+
i[\widetilde{\bm{H}}(t),\bm{K}(t)]^{\star}) \bm{\Lambda}(t)\}\bigr]\nonumber\\
&~-\eta\int_0^T \mathrm{d}t~|\pi(t)|^2/f(t),
\label{Y-tilde}
\end{eqnarray}
becomes maximal, where $\mathcal{Y}= \text{Re} \bigl[\text{Tr}[{\bm{\chi}^I}^{\dag}(T)\bm{E}_{00}]\bigr]$
and $\bm{\Lambda}(t)$ ($\eta$) is an operator (scalar) Lagrange multiplier. The last term in
Eq.~(\ref{Y-tilde}) describes an ``energy" constraint \cite{palao-kosloff}, in which
$f(t)$ is a shape function for switching the control field on and
off. In order to find the optimal field, we vary $\pi$, $\bm{\Lambda}$, and $a^I_{im}$,
and set $\delta\widetilde{\mathcal{Y}}=0$. By variation of the operator Lagrange multiplier $\bm{\Lambda}$,
we obtain the original dynamical equation [Eq.~(\ref{open-EQ})],
and variation of $\pi$ yields
\begin{eqnarray}
&\pi(t)=-\dfrac{f(t)}{2\eta}\text{Im}[\text{Tr}\left(\left[ \bm{\mu},\bm{K}(t)]^{\star}\bm{\Lambda}(t)\right)\right].
\label{pulse-eq}
\end{eqnarray}
 This equation implies that the knowledge of $\bm{K}(t)$ and $\bm{\Lambda}(t)$ is
necessary to specify the optimal control field. The superoperator $\bm{K}(t)$ can be
constructed by process estimation techniques. To obtain the dynamics of $\bm{\Lambda}(t)$ we vary
$a^I_{im}$, which in turn leads to variations of $\bm{\chi}^I$ and $\bm{K}$. Thus, the Lagrange
operator satisfies
\begin{eqnarray}
&-i\left[\bm{K}\dfrac{d\bm{\Lambda}}{dt}\right]_{imim}=\nonumber\\
&\sum_{njl}\Lambda_{ln}\widetilde{H}_{nimj}K_{imjn} -
\Lambda_{nm}\widetilde{H}_{mjli}\overline{K}_{jlim}.
\label{Lambda-dot}
\end{eqnarray}
Equations~(\ref{pulse-eq}) and (\ref{Lambda-dot}), in principle, can be
solved iteratively by the Krotov method \cite{palao-kosloff,krotov} to find the optimal
control field $\pi$ for decoherence suppression. That is, one can
effectively preserve coherence in dynamics of an open quantum system by applying external pulses
to decouple it from the environment. This could provide an alternative method for
an effective dynamical decoupling \cite{dyn-dec} in the language of
process matrix evolution. One can devise a learning decoherence control strategy by
estimating $\bm{K}(t)$, via certain QPT schemes on subensembles of identical systems, after each application
of the optimal control field in a given time $t$. The information learned from the
estimation is used through Eqs.~(\ref{Y-tilde})--(\ref{Lambda-dot}) for a second round to
find a new optimal $\pi$. This procedure can be repeated to enhance
the decoherence suppression task.

\textit{Conclusion and outlook.}---We have developed an alternative framework for monitoring and
controlling dynamics of open quantum systems, and have derived a
dynamical equation for the time variation of process matrices. This
nonperturbative approach can be applied to non-Markovian systems and
systems or devices strongly interacting with their embedding
environment. In addition, we have shown how the information gathered
via partial process tomography schemes can be used to efficiently
identify unknown parameters of
certain classes of local Hamiltonians in short-time scales. 
Furthermore, we have proposed an
optimal quantum control approach for the dynamics of open quantum
systems. Specifically, we have suggested how this mechanism can be
used for a generic decoherence suppression.

The approach presented here can be used for exploring new ways for
dynamical open-loop/learning control of Hamiltonian systems
\cite{JM}. One can utilize continuous weak measurements \cite{weak}
for process tomography to develop a real-time dynamical
closed-loop control for a quantum system. Our
Hamiltonian identification scheme could be utilized for efficient
verification of certain correlated errors for quantum computers and
quantum communication networks \cite{nielsen-book}. Using our dynamical approach,
one could explore the existence of certain symmetries in system-bath couplings
which would lead to noiseless subspaces and subsystems. The dynamical equation
developed here can also be applied to studying the energy transfer in multichoromophoric
complexes in the non-Markovian and/or strong interaction regimes \cite{mohseni-fmo}.


\textit{Acknowledgments.}---We thank A. Aspuru-Guzik, 
G. M. D'Ariano, N. Khaneja, P. G. Kwiat, D. A. Lidar, S. Lloyd, and B. C. Sanders
for helpful discussions. This work was supported by Faculty of Arts and Sciences of Harvard University, the USC Center for Quantum
Information Science and Technology, NSERC, \textit{i}CORE, MITACS, and PIMS.


\end{document}